\documentclass[preprint,12pt]{elsarticle}
\usepackage{float,subfigure,amsmath,graphicx,tabularx,multicol}
\journal{Organic Electronics}

\begin{document}

\begin{frontmatter}

\title{Living Wires --- Effects of Size and Coating of Gold Nanoparticles in Altering the Electrical Properties of \emph{Physarum polycephalum} and Lettuce Seedlings}
%\author[label1,label1,label1,label2,label2]{Nina Gizzie*, Richard Mayne, Andrew Adamatzky, Shlomo Yitzchaik \& Md Ikbal}
\author[label1]{Nina Gizzie\corref{1}}
%\ead{Nina.Gizzie@Gmail.com}
\author[label1]{Richard Mayne}
\author[label2]{Shlomo Yitzchaik}
\author[label2]{Muhamad Ikbal}
\author[label1]{Andrew Adamatzky}

\cortext[1]{Corresponding author, Nina.Gizzie@gmail.com}

\address[label1]{Unconventional Computing Group, University of the West of England, Bristol, UK}
\address[label2]{Institute of Chemistry, The Hebrew University of Jerusalem, Jerusalem, Israel}

\begin{abstract}
The manipulation of biological substrates is becoming more popular route towards generating novel computing devices. \emph{Physarum polycephalum} is used as a model organism in biocomputing because it can create `wires' for use in hybrid circuits; programmable growth by manipulation through external stimuli and the ability withstanding a current and its tolerance to hybridisation with a variety of nano/microparticles. Lettuce seedlings have also had previous interest invested in them for generating plant wires, although currently there is little information as to their suitability for such applications. In this study both \emph{P. polycephalum} and Lettuce seedlings were hybridised with gold nanoparticles --- functionalised and unfunctionalised --- to explore their uptake, toxicological effects and, crucially, any alterations in electrical properties they bestow upon the organisms. Using various microscopy techniques it was shown that \emph{P. polycephalum} and lettuce seedlings are able to internalize nanoparticles and assemble them \emph{in vivo}, however some toxicological effects were observed. The electrical resistance of both lettuce seedlings and \emph{P. polycephalum} was found to decrease, the most significant reduction being with lettuce seedlings whose resistance reduced from 3M$\Omega$s to 0.5M$\Omega$s. We conclude that gold is a suitable nanomaterial for biohybridisation specifically in creating conductive pathways for more efficient biological wires in self-growing hybrid circuitry.\\
\end{abstract}

\end{frontmatter}

%% To do 06/10/15
% Abstract
% Tidy discussion
% Fig. 2
% State concentrations of particles

\section{Introduction}

This study was designed to investigate the enhancement in functionality of hybrid artificial-living substrates via integration of nanoscale metallic components towards the generation of unconventional computing devices. Previous work in the field has shown promise in implementing computing devices utilising living organisms to create self-healing circuits loaded with nano- and microstructures: one of the major routes towards this is by utilising slime mould \emph{Physarum polycephalum} as a model organism \cite{Adamatzky2013, adamatzky2013towards, Mayne2015}. 

\emph{P. polycephalum} is a large multinucleate single celled organism which can grow up to 30~cm$^2$ and has the ability to migrate towards food sources by forming a dynamic fan-shaped advancing anterior margin which recedes into an interlinking network of protoplasmic tubular structures \cite{Kessler1982}. These tubes have been shown to use contractile proteins in the ectoplasm (peripheral layer sitting circumferentially about the diameter of tubes) to instigate monorhythmic oscillation of the endoplasm (hydrodynamic core of tubes) allowing the organism to distribute nutrients and provide motive force \cite{Kessler1982}. \emph{P. polycephalum} is a malleable unconventional computing substrate as it is able to display a range of apparently intelligent behaviours and expressions of natural computing:  for example it is able to optimize nutrient harvesting networks (route planning), solve maze patterns when guided by chemotactic gradients and approximate Voronoi diagrams 
\cite{Nakagaki2001, adamatzky2008towards,Adamatzky2009,Costello2014,Shirakawa2009, adamatzky2011approximating}. \emph{P. polycephalum} is also suggested to perform like a neuron with regards to its electrical excitability \cite{Gale2013} as it produces and responds to electrical signals. We refer the reader to a summary of plasmodial electrical activity by Mayne \& Adamatzky \cite{Mayne2015}. With regards to \emph{P. polycephalum}'s ability to withstand integration into conventional circuitry, it is able to comfortably withstand currents of 50~$\mu$A \cite{Tsuda2011}.
%: although it has a high electrical resistance it does somewhat permit the flow of current \cite{Mayne2015}.
To demonstrate the efficacy of hybrid artificial-\emph{P. polycephalum} circuitry, individual plasmodial tubes may be readily integrated into microcontrollers and FPGAs to create basic sensing devices \cite{Mayne2015b}. 

More recently it has been shown that shown that \emph{P. polycephalum} can be loaded with microparticles\cite{Cifarelli2014} and nanoparticles \cite{Mayne2011,Mayne2015} of a wide range sizes and compositions; these studies demonstrate the organism's ability to transport and redistribute internalised particles through its protoplasmic tubes without loss of its observable natural biological activity and functions (streaming dynamics, electrical phenomena etc.). Other, less conventional uses for this hybridised plasmodia may be realised through the use of different particle varieties, such as biomorphic conductive pathways, generation of nanostructures and easier manipulation of \emph{P. polycephalum} behaviours, as well as mapping nanoelectrodes on functionalised surfaces \cite{Cifarelli2014}. \emph{P. polycephalum} however does have its shortfalls as a computing substrate: growth is sporadic and difficult to control, environmental factors play a significant role in successful development and the tubes themselves are delicate. 

This study details an investigation into computing applications of \emph{P. polycephalum} hybridised with a range of gold nanoparticles: gold is a desirable substance for such purposes as it is highly conductive, unreactive and easily functionalised, but has not been previously utilised with slime mould. We also propose to investigate a second biological substrate in order to assess the suitability of gold nanoparticles for this application in other tissues. The second variety of substrate used was lettuce seedlings (\emph{Lactuca sativa}), which was chosen due to previous interest in generating plant wires \cite{Adamatzky2014}. Although there is little information on plant roots and their suitability within a hybrid computer, the aforementioned study has shown that lettuce is able to withstand a current and produces a similar resistance to that of \emph{P. polycephalum}. As plants are a far more robust and less temperamental substrate --- e.g. it has no aversion to natural light --- the unconventional computing applications of such a substrate may be broader. 

The following investigation assesses the uptake, toxicological effects and bestowed electrical properties of a variety of gold nanoparticles in these organisms via electrophysiological measurements and various microscopic techniques, to assess the suitability for generating self-growing hybrid artificial circuitry.

\section{Methods}
\subsection{Culture and Nanoparticle Hybridisation}

Plasmodia of \emph{P. polycephalum} (strain HU554 $\times$ HU560) were cultivated and treated with nanoparticle suspension as described in Mayne\cite{Mayne2011};  \emph{P. polycephalum} was cultured on paper towels moistened with distilled water in a plastic container. Oat flakes were dispensed around the initial colony and the container was sealed and placed in the dark at room temperature. Healthy organisms were then subcultured by taking a \emph{P. polycephalum}-colonised oat flake and placing it onto the centre of a 2\% non-nutrient agar plate in a 9~cm plastic Petri dish. To facilitate growth oat flakes were placed in periphery of the dish and left to migrate in the dark for up to 48 hours. \emph{P. polycephalum} plasmodia receiving nanoparticle treatment were prepared using a feeding method; 25~$\mu$l of nanoparticle suspension was dispensed directly onto the colony. A small moat was cut into the agar around colonies to prevent the NP solution diffusing through the agar. Colonies were left to migrate to oat flakes for up to 48 hours.

Lettuce seedlings (`red oakleef' \emph{Lactuca sativa}) were germinated as follows: seeds were placed in a Petri dish on top of filter paper moistened with deionised water. 1~ml of water was applied to the filter paper initially and then 500~$\mu$l each day until the seeds started germinating. Seedlings that did not germinate after 48 hours were not utilised. After 48 hours 25~$\mu$l of nanoparticle suspension was dispensed on top of each lettuce seedling, which were left to grow for up to 72 hours until the seedlings reached between 10--15~mm.

Three types of gold Nanoparticles were used:
\begin{enumerate}
\item 5~nm diameter, stabilised in 0.1nM PBS (Sigma Aldrich, UK).
\item 200~nm diameter, stabilised in 0.1nM citrate buffer (Sigma Aldrich, UK). %% Buffer str?
\item 2~nm Spiropyran-modified gold nanoparticles.
\end{enumerate}
Both unfunctionalised 5~nm and 200~nm nanoparticles were chosen for their minimal reactivity (in comparison with some other conductive nanomaterials) and their sizes: the 5~nm variety had an approximately equal diameter to the functionalised particle's hydrodynamic diameter and the 200nm variety were chosen due to previous successes in the field with particles of this size~\cite{Mayne2015}. The 2~nm spiropyran-modified-modified particles were chosen for the photochromic properties. The molecular switch compound reversibly converts from the bulky non-polar spiropyran form into a planar and polar merocyanine, an aggregative protein, when exposed to UV light, and hence bestows the suspension with a highly desirable property --- variable resistance based on the degree of particle aggregation~\cite{Shiraishi2014}. Hence one aim of this study was to assess whether these functionalised nanoparticles retained their properties \emph{in vivo}.

\subsection{Scanning Electron Microscope and Microanalysis}

\emph{P. polycephalum} tube samples were taken 3~cm away from the site of inoculation, including their agar base, and were air dried overnight. Lettuce seedlings were similarly prepared. Samples were mounted on SEM stubs with double sided carbon tabs and were viewed in a Philips XL30 environmental scanning electron microscope. Samples were viewed under high vacuum. Energy-dispersive x-ray spectroscopy (EDX) was used for all samples using an Oxford Instruments Link system.
	 
\subsection{Preparation of Semi-thin Sections}

\emph{P. polycephalum} samples were fixed in 2.5\% glutaraldehyde, 2.5\% para\-formaldehyde in 0.1M pH 7.2 potassium phosphate buffer for 15 minutes, which were then rinsed in the same buffer once per hour for 3 hours. Lettuce samples were fixed in 4\% parafomaldehyde with 0.1\% DMSO in 0.1M pH 6.85 potassium phosphate buffer and left to fix overnight at 2--4$^o$C. Samples were then rinsed as above.
Both \emph{P. polycephalum} and lettuce samples were then dehydrated using a graduated series of ethanol ($30\rightarrow 50\rightarrow 70\rightarrow 90\rightarrow [100\times2]$) with 15 minute intervals. This was followed by infiltration with 100\% LR white hard resin (Agar Scientific, UK) for two 1 hour changes followed by one overnight on a rotator. Tissue pieces were embedded in gelatin capsules and thermally polymerised for 24 hours at 60$^o$C$\pm 2$. Sections were cut using a Reichert-Jung Ultracut E ultramicrotome. Semi-thin sections (1~$\mu$m) were taken for staining with toluidine blue for light microscopy and uranyl acetate (15 minutes) and lead citrate (5 minutes) for SEM analysis. 

\subsection{Histology} 

\emph{P. polycephalum} samples were fixed in 4\% paraformaldehyde in 0.1M pH 7.2 potassium phosphate buffer for 90 minutes before being rinsed 3 times in the same buffer. Lettuce samples were fixed in 4\% parafomaldehyde in pH 6.85 potassium phosphate buffer with 0.1\% DMSO and left to fix for 2 hours, followed by 3 rinses. 

All samples were dehydrated using increasing concentrations of ethanol ($70\rightarrow 95\rightarrow [100\times2]$, 15 minute changes). Samples then underwent clearing in 3 15 minute changes of limonene. This was followed by infiltration with 50:50 mix of limonene and paraffin wax followed by 100\% wax. Samples were embedded in transverse orientation and cut at 4~$\mu$m on a Leica RM 2235 rotary microtome. Sections were stained using haematoxylin and eosin staining as per local protocols.

\subsection{Electrical Properties}

\begin{figure}[!tbp]
\centering
\subfigure[]{\includegraphics[width=0.9\textwidth]{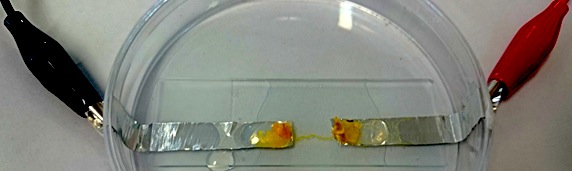}}
\subfigure[]{\includegraphics[width=0.9\textwidth]{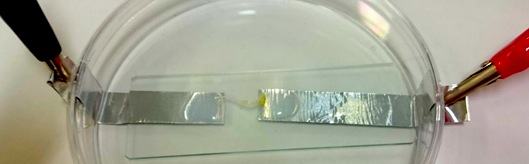}}
\caption{Set up for electrical measurements. (a)~10mm Physarum 'wire' grown between two agar islands overlying aluminum foil electrodes. (b)~Lettuce seedling placed upon two agar islands overlying aluminum foil electrodes }
\label{fig-electricalSetup}
\end{figure}

The test circuit for both lettuce seedlings and \emph{P. polycephalum} are shown in Fig. \ref{fig-electricalSetup}. Two 200~$\mu$l 2\% non-nutrient agar islands of were placed overlying two 6$\times$90~mm aluminium tape electrodes separated by a 10~mm gap. 

\emph{P. polycephalum} samples were homogenised with a scalpel and placed on one island, a fresh oat flake was placed on the other to encourage directional growth. Plates were sealed with paraffin film. \emph{P. polycephalum} was left to migrate for up to 48 hours creating a wire between the two electrodes. 

10--15~mm lettuce seedlings were placed on top of similar agar islands for measurement.

 %% Do we need the pic of the datalogger? %% Original pic 1

Voltage measurements were made with a Picotech ADC-24 Datalogger (Pico, UK). Resistance was measured using a Keithley SMU 2400 (Keithley, USA) using a 1~$\mu$A test current. \emph{P. polycephalum} were measured for 2000 seconds to show oscillations; those that didn't oscillate were discounted. Lettuce seedlings were measured for 500 seconds as their properties were found to be comparatively stable. 

All data were analysed using IBM SPSS version 20. Normality of data and Variances of Homogeneity were assessed. ANOVA one way was used where data was normal and variances were met. Welch test was used when normality was met but variances were not. Kruskall-Wallis was used when normality was not met. All tests were used with a significance level of p=0.05. For analysis of slime mould oscillatory activity, data were Fast Fourier Transformed (FFT) with Sigview software (V2.7.1). 

\section{Results}

\subsection{Macroscopic Observations and Nanoparticle Toxicity}
\begin{figure}[!tbp]
\centering
\subfigure[]{\includegraphics[width=0.49\textwidth]{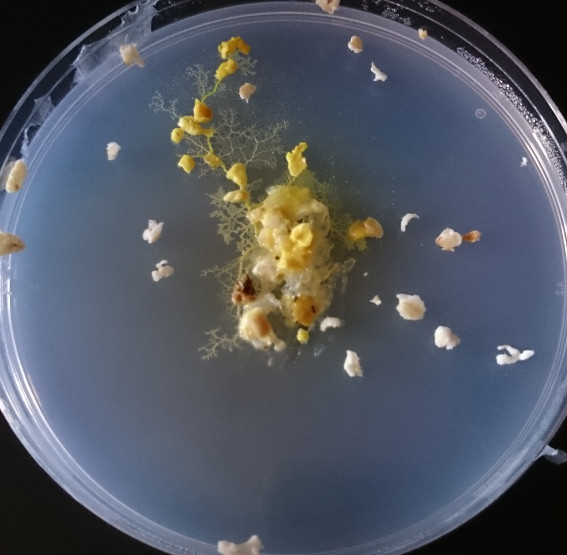}}
\subfigure[]{\includegraphics[width=0.49\textwidth]{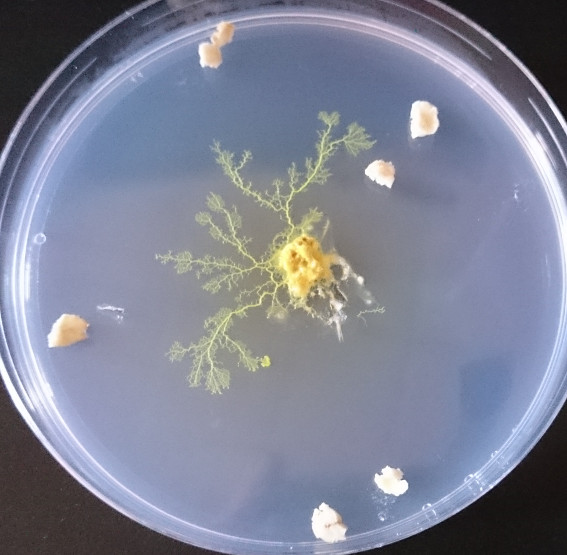}}
\subfigure[]{\includegraphics[width=0.49\textwidth]{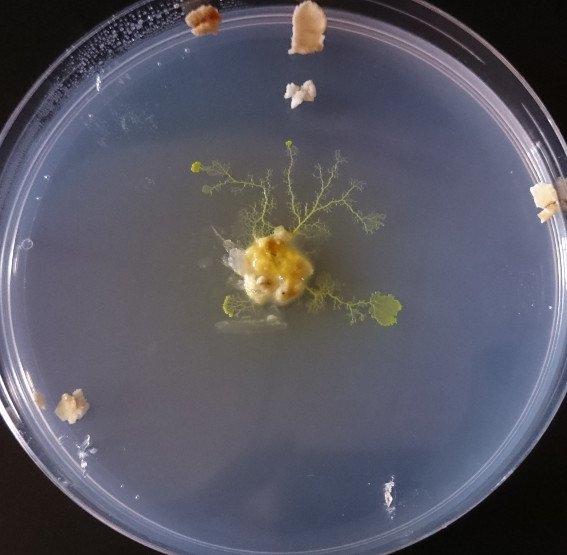}}
\subfigure[]{\includegraphics[width=0.49\textwidth]{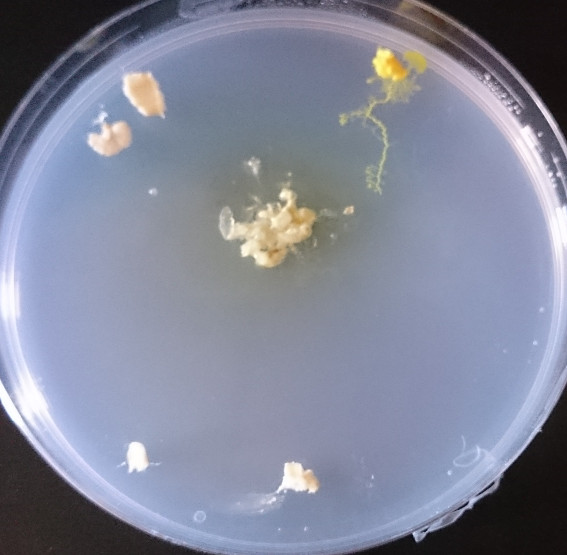}}
\caption{Macroscopic appearance of experimental \emph{P. polycephalum} 24 hours after inoculation. (a) Control; (b) Unfunctionalised 5~nm; c) Unfunctionalised 200~nm; d) Spiropyran-modified.}
\label{fig-phymacro}
\end{figure}

\emph{P. polycephalum} appeared to tolerate the presence of all nanoparticle varieties despite discolouration and morphological differences in some experimental plasmodia (Fig.~\ref{fig-phymacro}). With both varieties of unfunctionalised particle a somewhat reduced growth rate was observed with when compared to controls, although the tubes are still of a similar appearance. \emph{P. polycephalum} treated with Spiropyran-modified nanoparticles (Fig.~\ref{fig-phymacro}d) was however significantly emaciated and was observed to rapidly vacate its inoculation point. 

 %% Original pic 3 %% Not finished

\begin{figure}[!tbp]
\centering
\subfigure[]{\includegraphics[width=0.49\textwidth]{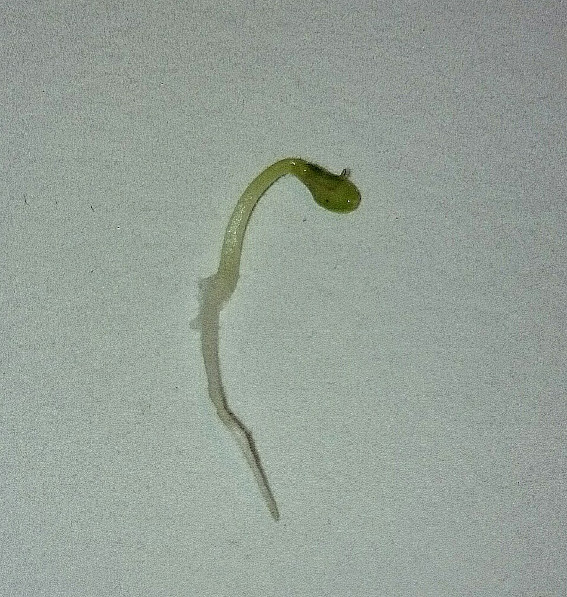}}
\subfigure[]{\includegraphics[width=0.49\textwidth]{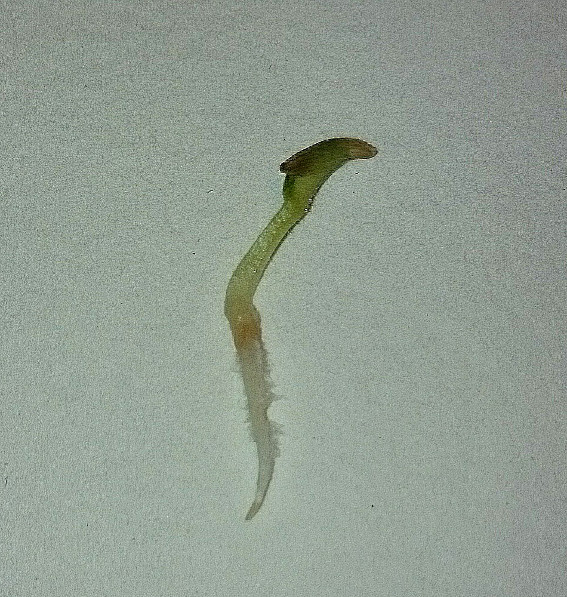}}
\subfigure[]{\includegraphics[width=0.49\textwidth]{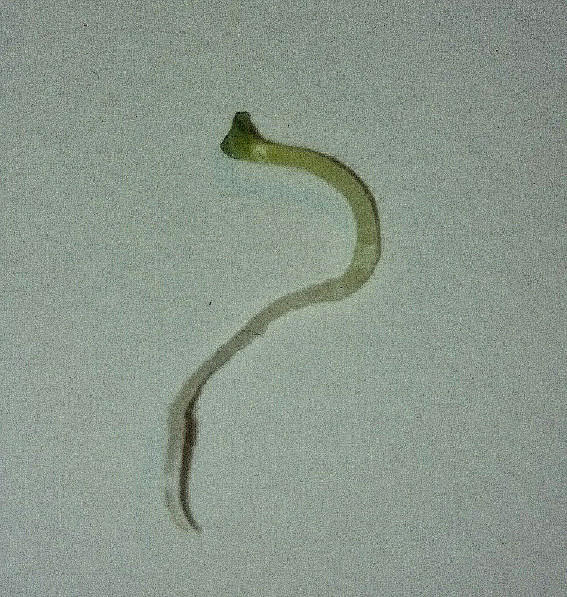}}
\subfigure[]{\includegraphics[width=0.49\textwidth]{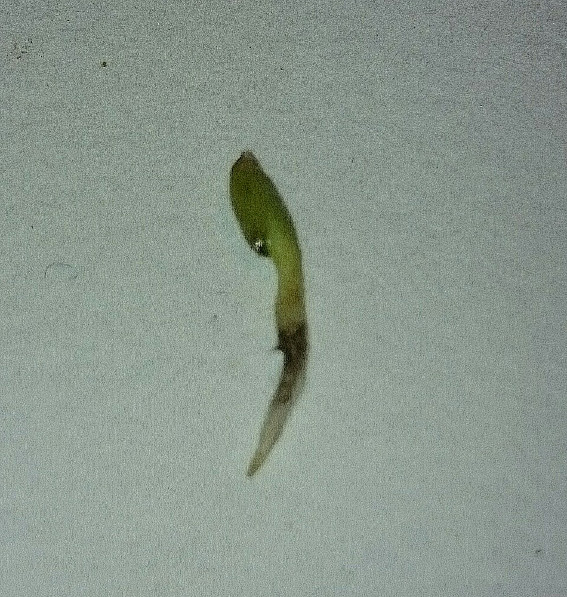}}
\caption{Macroscopic appearance of Lettuce seedlings 48 hours after exposure to nanoparticles. Arrowheads indicate discolouration. (a) Control; b) unfunctionalised, 5nm; (c) unfunctionalised, 200nm; (d) Spiropyran-modified. }
\label{fig-LetMacro}
\end{figure}

Lettuce seedlings appeared to tolerate the presence of all nanoparticles varieties (Fig. \ref{fig-LetMacro}). Lettuce seedlings treated with unfunctionalised 5~nm and 200~nm nanoparticles did not show any toxic properties. The only indication that the unfunctionalised nanoparticle suspensions had affected the seedlings was a slight discolouration in the stem (Fig. \ref{fig-LetMacro}b). Lettuce seedlings treated with Spiropyran-modified nanoparticles did show some reduction in growth rate and germination frequency.

 %% Original pic4  %%  Mention arrows

In the more inhospitable environment of the electrical testing set up, nanoparticle-treated \emph{P. polycephalum} and lettuce seedlings were found to have a higher propagation failure rate (approximately 4-fold for lettuce and 8-fold for \emph{P. polycephalum}).

\subsection{Microscopy and Microanalysis}

\subsubsection{{P. polycephalum}}
\begin{figure}[!tbp]
\centering
\subfigure[]{\includegraphics[width=0.49\textwidth]{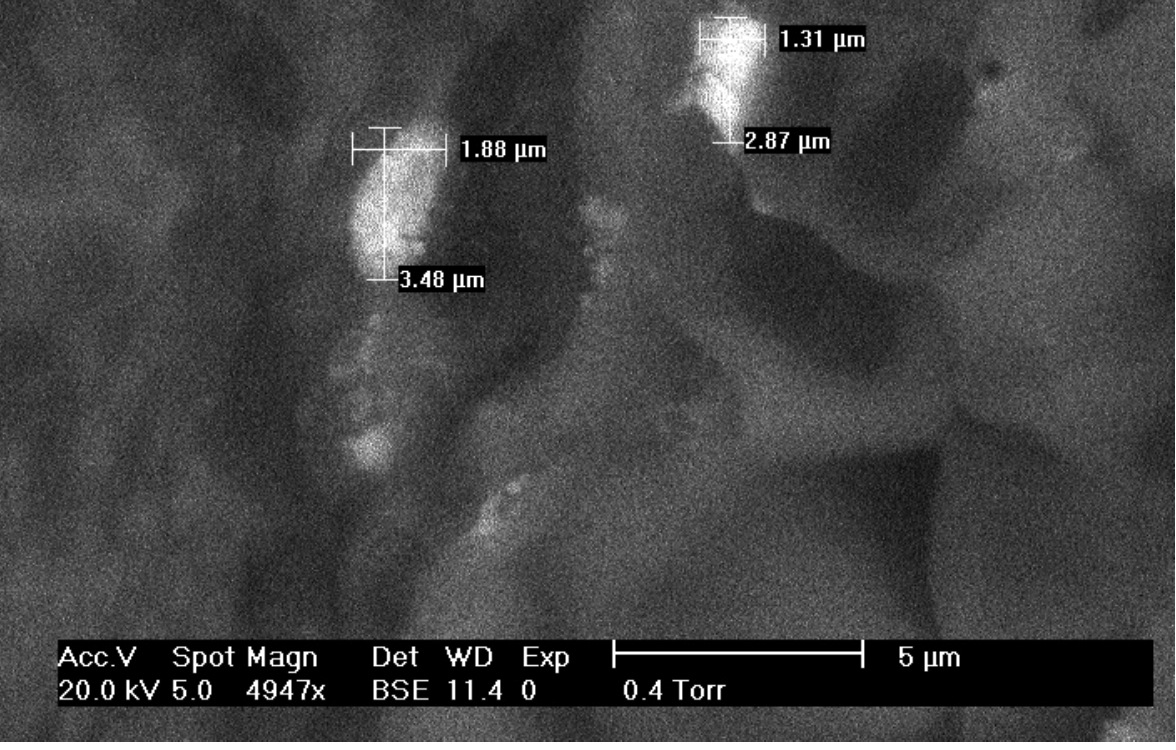}}
\subfigure[]{\includegraphics[width=0.49\textwidth]{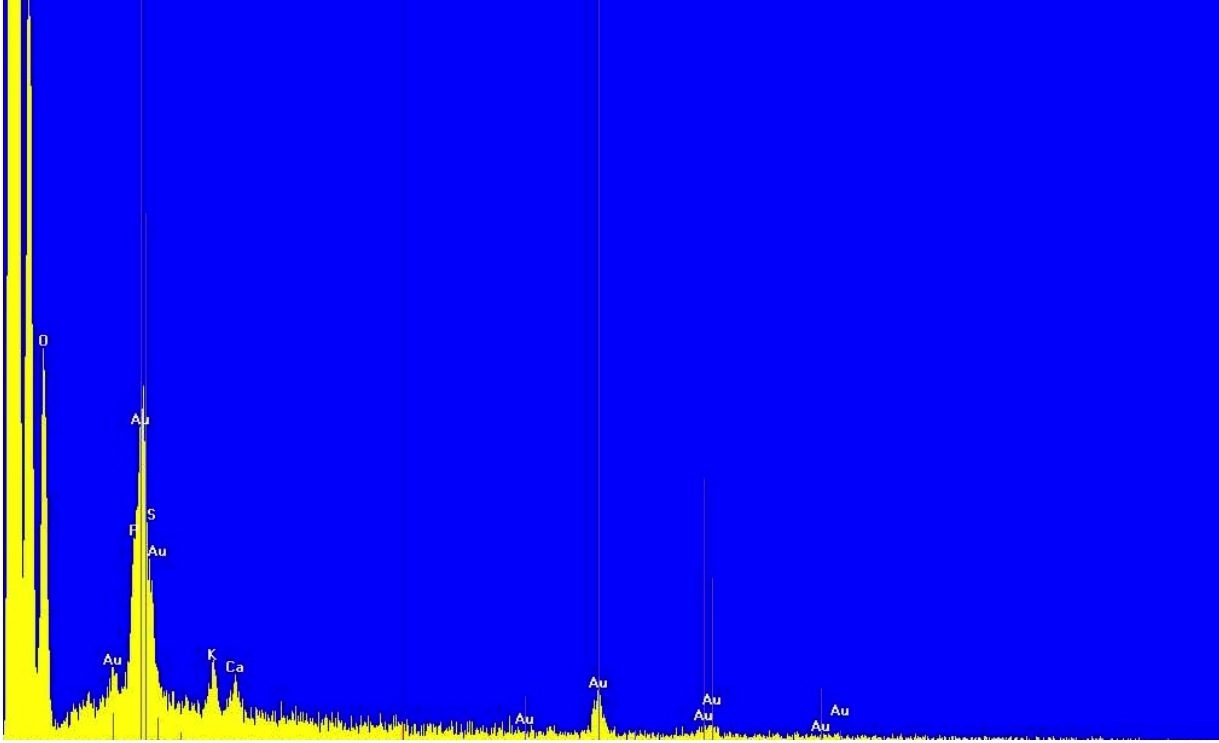}}
\subfigure[]{\includegraphics[width=0.49\textwidth]{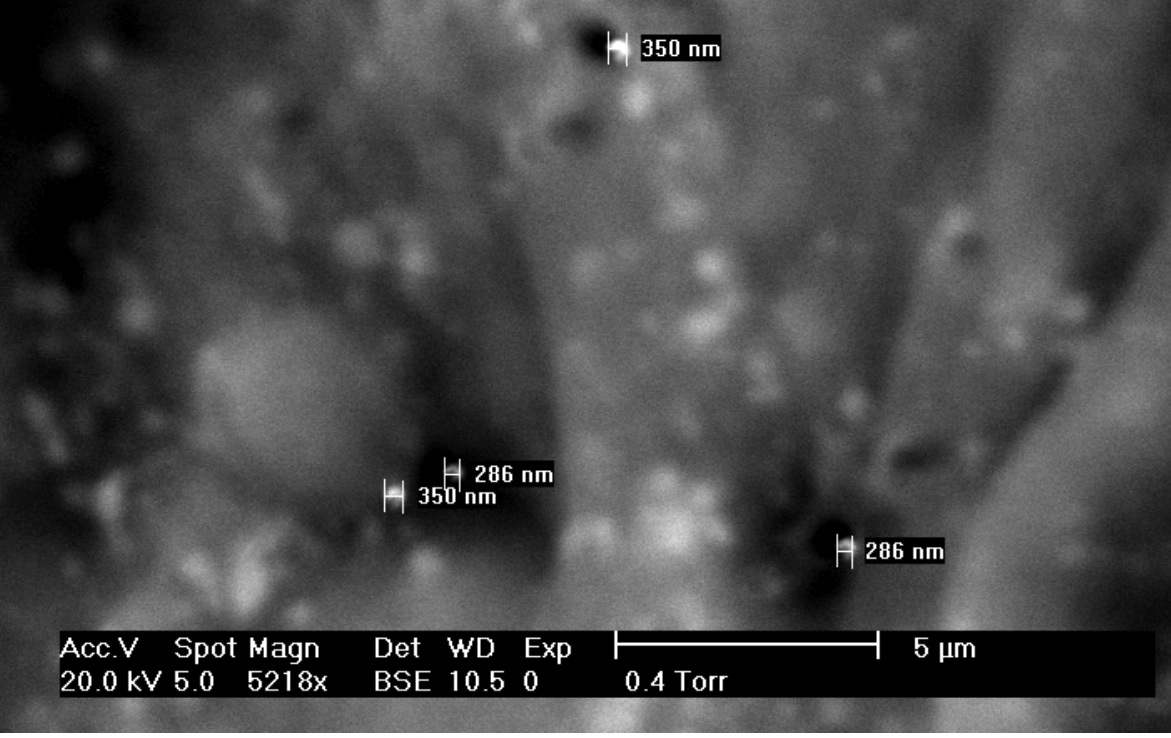}}
\subfigure[]{\includegraphics[width=0.49\textwidth]{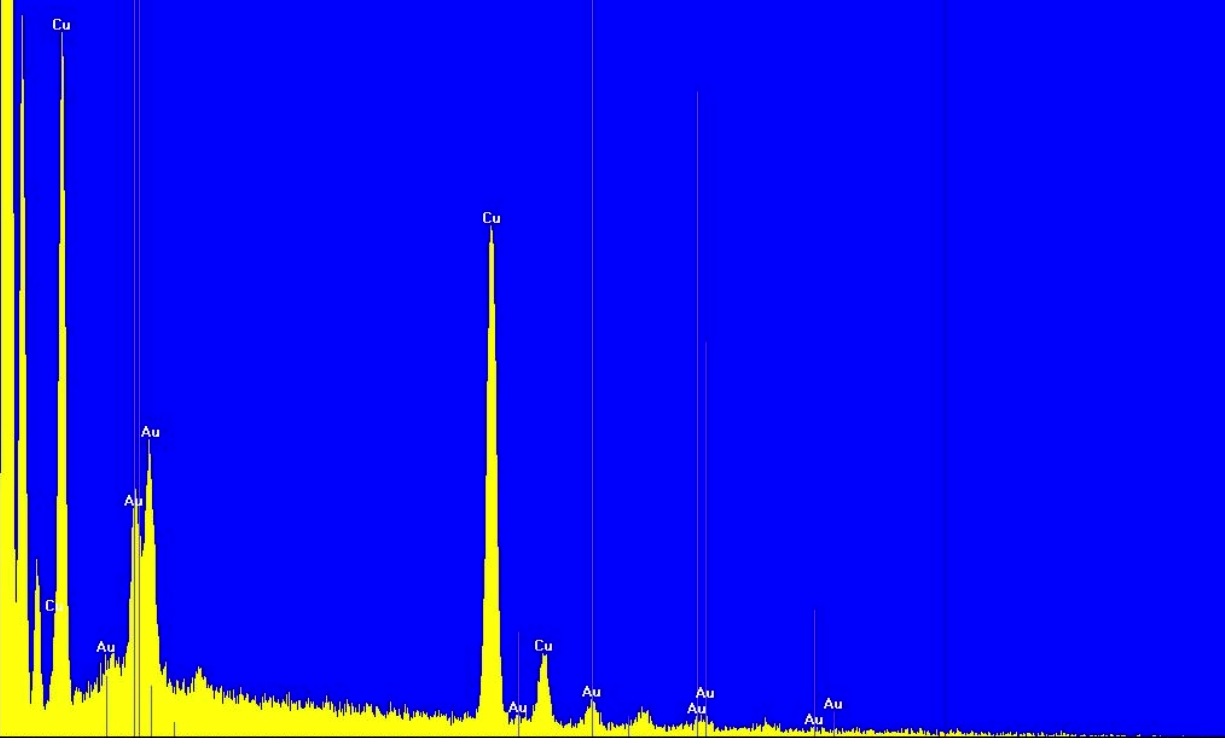}}
\subfigure[]{\includegraphics[width=0.49\textwidth]{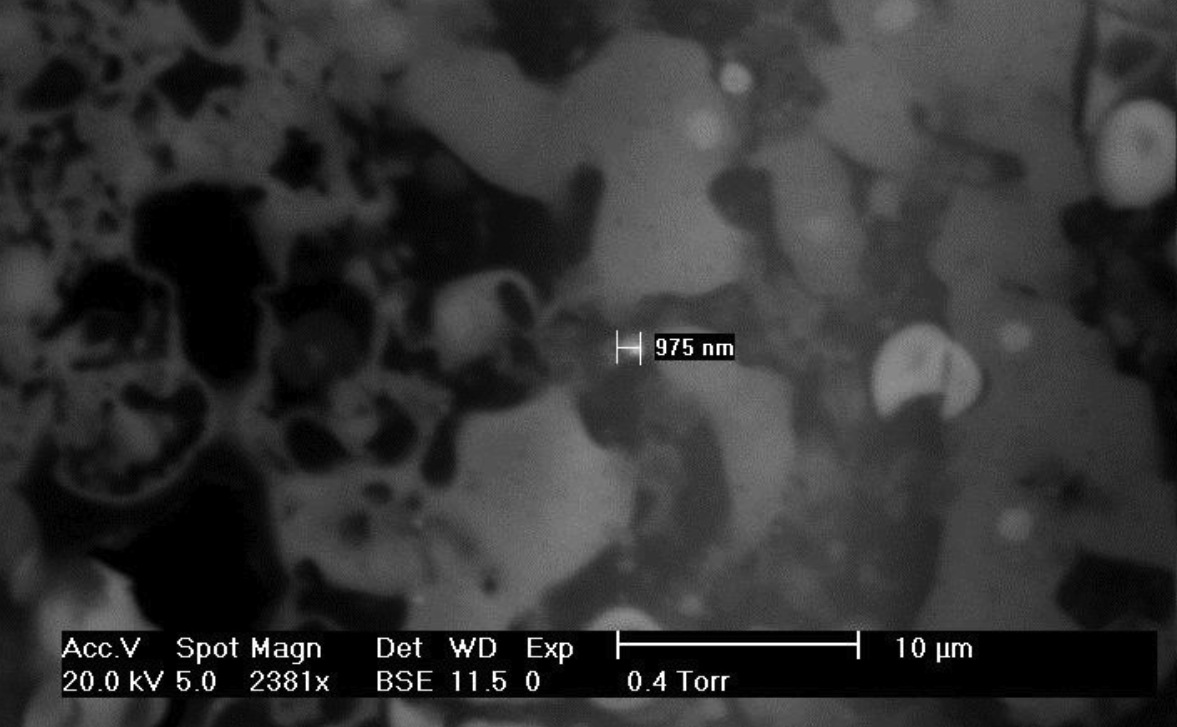}}
\subfigure[]{\includegraphics[width=0.49\textwidth]{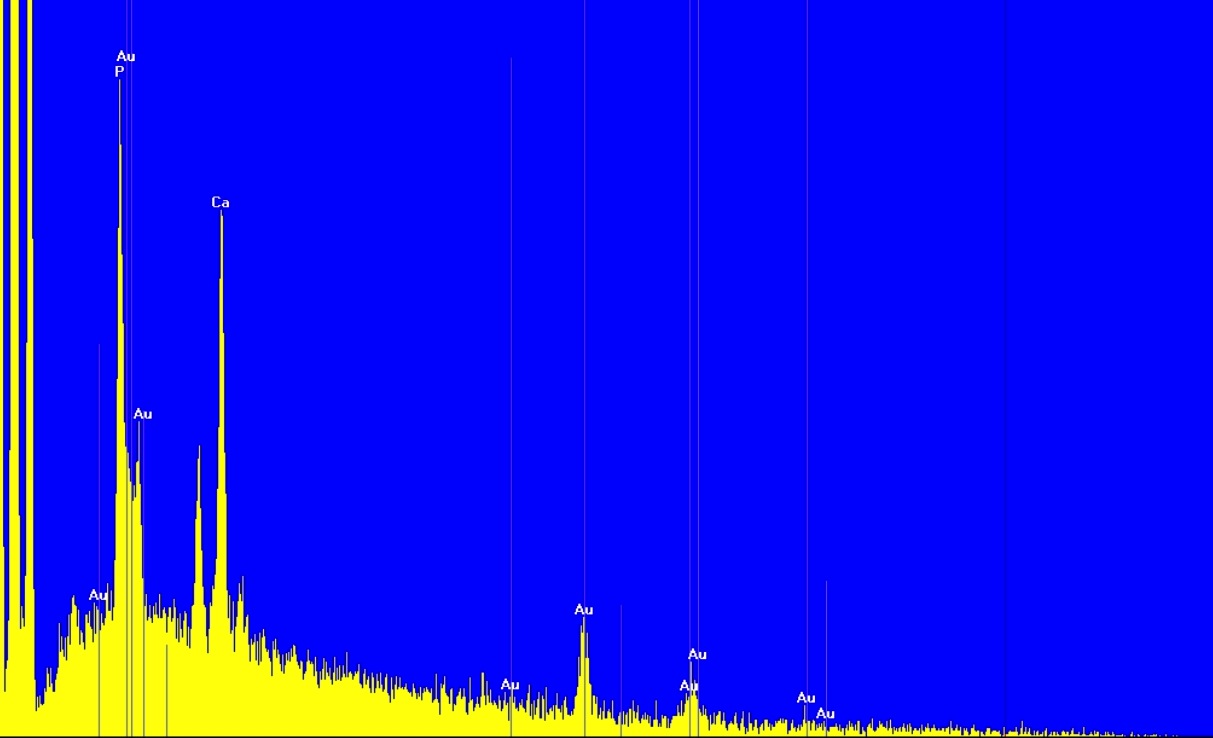}}
\caption{Scanning electron micrographs and corresponding EDX analyses of \emph{P. polycephalum} treated with various gold nanoparticles. Bright regions in (ace) correspond to areas of high elemental number content. Gold is confirmed in each section in EDX spectra (bdf). (a) Surface view of plasmodium treated with Spiropyran-modified particles and (b) its EDX spectrum. (c) Surface view of plasmodium treated with unfunctionalised 200~nm particles and (d) its EDX spectrum. (e) Transverse section through a plasmodial tube treated with unfunctionalised 200~nm particles and (f) its EDX spectrum.}
\label{fig-PhySem}
\end{figure}

Using Scanning electron microscopy (SEM) via back scattered electron detection, areas of high atomic number content were identified (represented by bright spots) within and on the surface of \emph{P. polycephalum} tubes (Figs. \ref{fig-PhySem}a and \ref{fig-PhySem}c); gold presence was confirmed by EDX analysis within both Spiropyran-modified (Fig. \ref{fig-PhySem}b) and unfunctionalised 200~nm (Fig. \ref{fig-PhySem}d) treated tubes. All gold objects in spiropyran-modified nanoparticle-treated plasmodia were found to be micrometre-scale, i.e. significantly larger than the size of the individual nanoparticles (Fig. \ref{fig-PhySem}a). The unfunctionalised 200~nm particles were present in deposits between 200-400~nm. Both unfunctionalised 200~nm and Spiropyran-modified particles were only found in desiccated tubes or at the end of trails. No unfunctionalised 5~nm were seen using SEM.

\begin{figure}[!tbp]
\centering
\subfigure[]{\includegraphics[width=0.49\textwidth]{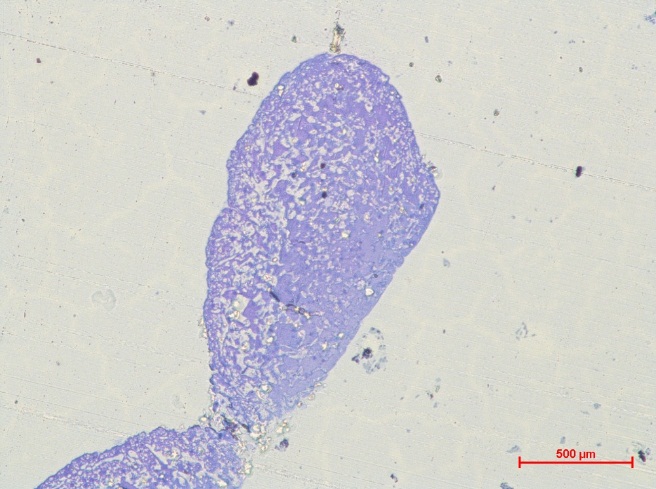}}
\subfigure[]{\includegraphics[width=0.49\textwidth]{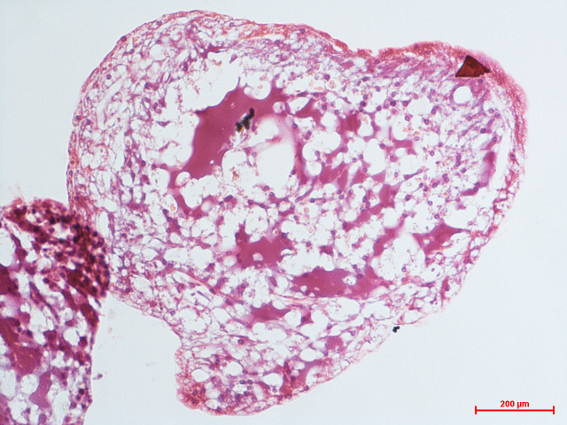}}
\caption{Light micrographs of \emph{P. polycephalum} histological sections, transverse orientation. (a) Toluidine blue stain, 1~$\mu$m section. (b) Haemotoxylin and eosin stain, 4~$\mu$m section.}
\label{fig-PhyHisto}
\end{figure}

Indications of gold presence were not found in \emph{P. polycephalum} 4~$\mu$m or 1~$\mu$m histological sections (Fig. \ref{fig-PhyHisto}). Figure \ref{fig-PhySem}e,f, however, demonstrates the presence of gold in SEM 1~$\mu$m transverse sections of \emph{P. polycephalum} treated with unfunctionalised 200~nm particles. 

 %% Original pic 5[ab],6[cd],8[ef] %% Indicate peaks % formerly `SemPhySp'

 %% Original pic 7 %% WHAT WERE THESE TREATMENTS?

Indications of gold presence were not found in \emph{P. polycephalum} 4~$\mu$m or 1~$\mu$m histological sections (Fig.~\ref{fig-PhyHisto}). Figure~\ref{fig-PhySem}e,f however demonstrates the presence of gold in SEM 1~$\mu$m transverse sections of \emph{P. polycephalum} treated with unfunctionalised  200nm particles.

\subsubsection{Lettuce seedlings}

\begin{figure}[!tbp]
\centering
\subfigure[]{\includegraphics[width=0.49\textwidth]{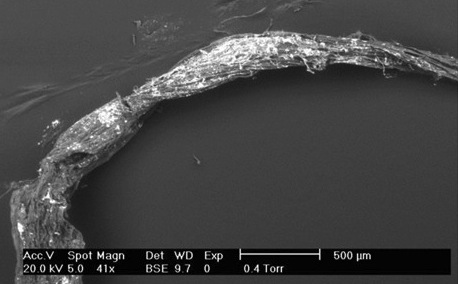}}
\subfigure[]{\includegraphics[width=0.49\textwidth]{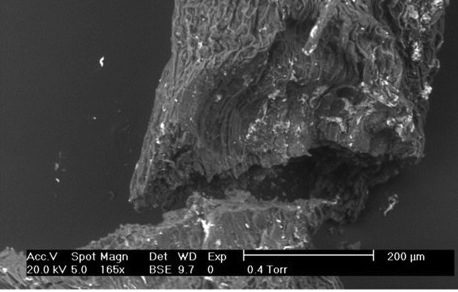}}
\subfigure[]{\includegraphics[width=0.49\textwidth]{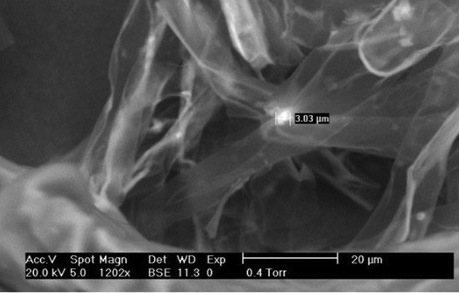}}
\subfigure[]{\includegraphics[width=0.49\textwidth]{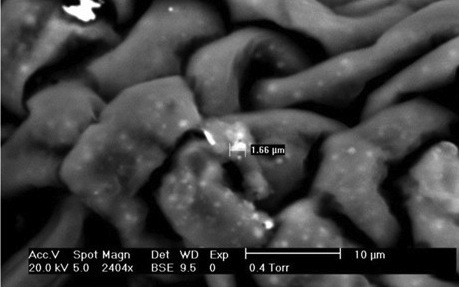}}
\caption{Scanning electron micrographs of lettuce seedlings treated with various nanoparticle suspensions. Bright regions in correspond to areas of high elemental number content. All samples were confirmed to contain gold via EDX (data not shown) (a) Dehydrated seedling treated with Spiropyran-modified nanoparticles. (b) Interior of sample in $\langle$A$\rangle$. (c) Dehydrated seedling treated with unfunctionalised 5~nm particles. (d) Dehydrated seedling treated with unfunctionalised 200~nm particles.}
\label{fig-LetSem}
\end{figure} %% Original pic 9[ab],10[cd] % Formerly "semletsp"

SEM back-scattered electron detection revealed the presence of high atomic number content areas in all samples treated with nanoparticle suspensions (Fig.~\ref{fig-LetSem}): EDX analysis confirmed the presence of gold within lettuce seedlings treated with spiropyran-modified-modified (Fig.~\ref{fig-LetSem}c), unfunctionalised 5~nm (Fig.~\ref{fig-LetSem}c) and 200~nm (Fig.~\ref{fig-LetSem}d). Spiropyran-modified nanoparticle-treated seedlings showed especially large areas of gold on the surface of the lettuce seedling (Fig.~\ref{fig-LetSem}a,b).

\begin{figure}[!tbp]
\centering
\subfigure[]{\includegraphics[width=0.49\textwidth]{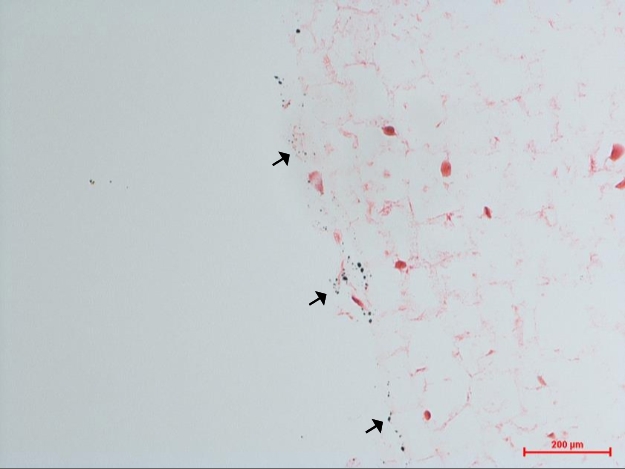}}
\subfigure[]{\includegraphics[width=0.49\textwidth]{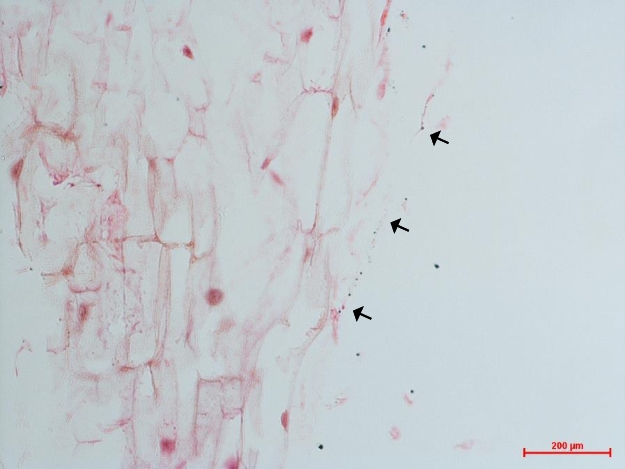}}
\caption{Light micrographs of lettuce seedling 4~$\mu$m histological sections treated with gold nanoparticles, haematoxylin and eosin stain. Arrows indicate black deposits in epidermis. (a) Seedling treated with unfuncctionalised 200~nm particles. (b) Seedling treated with unfuncitonalised 5~nm particles. Epidermal deposits are smaller than those in the sample depicted in $\langle$A$\rangle$.}
\label{fig-LetHisto}
\end{figure}

4~$\mu$m lettuce seedling sections showed the presence of large black deposits in the epidermis of samples treated with unfunctionalised 200~nm (within the stem) (Fig. \ref{fig-LetHisto}a) and unfunctionalised 5~nm (within the root) (Fig. \ref{fig-LetHisto}b). The deposits in seedlings treated with unfunctionalised 5~nm nanoparticles were smaller and distributed more sparsely than those of the unfunctionalised 200~nm. Due to their similarity in size and distribution to the deposits viewed via SEM, these were reasoned to be gold.

 %% Original pic 11,12

\subsection{Electrical Properties}
\subsubsection{Voltage}

\begin{figure}[!tbp]
\centering
\includegraphics[width=1\textwidth]{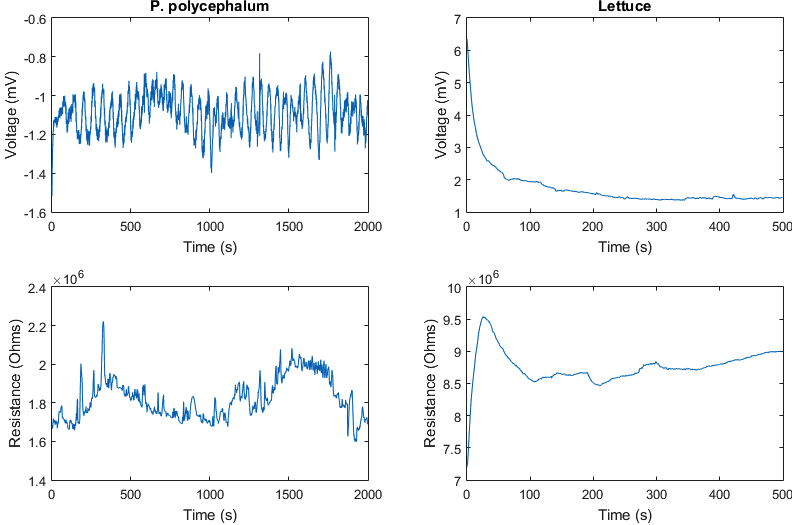}
\caption{Electrical characterisation of exemplar \emph{P. polycephalum} and lettuce seedling controls. \emph{P. polycephalum} readings taken over 2000 seconds; its electrical potential may be observed to oscillate rhythmically and so can the resistance, although to a lesser extent. Lettuce seedlings show exponential decay of electrical potential and an approximately corresponding increase in resistance.}
\label{fig-Traces}
\end{figure} %% Original pic 13,14 + 2 new ones

\begin{table}[!tbp]
\caption{Electrical potential measurement analysis for lettuce seedlings treated with gold nanoparticles. UF5/200: Unfunctionalised 5/200~nm varieties, SPM: spiropyran-modified.}
\centering
%\scriptsize
{\begin{tabularx}{\linewidth} {@{} p{2.5cm} c p{2cm} p{2cm} p{1.5cm} r p{1.0cm}@{}} 

%{@{} p{2.5cm} c p{2cm} p{2cm} p{1.5cm}p{1cm}r@{}}  
\hline
Treatment & Mean Potential (mV) & SD & SE & P value\\
\hline
Control & 1.9 & 2.87 & 0.70 & --\\
UF5 & 3.5 & 2.9 & 0.69 & 0.460\\ 
UF200 & 3.0 & 3.6 & 0.77 & 0.693\\
SPM & 4.0 & 3.2 & 0.97 & 0.342\\
\hline
\end{tabularx}}
\label{table-LetVol}
\end{table}

\begin{table}
\caption{Electrical potential measurement analysis for \emph{P. polycephalum} treated with gold nanoparticles. UF5/200: Unfunctionalised 5/200~nm varieties.}
\centering
\tiny
{\begin{tabularx}{\textwidth}{@{}lXcXXXcXXr@{}}
%\toprule
\hline
Treatment & Period (s) & Mean Potential (mV) & SD & SE & P Value & \multicolumn{1}{X}{Dominant Frequency (Hz$\times10^{-3}$)} & SD & SE & P Value\\
%\colrule
\hline
Control & 193.51 & 0.52 & 0.37 & 0.10 & -- & 5.17 & 0.006 & 0.001 & --\\
UF5 & 232.58 & 0.72 & 0.94 & 0.31 & 0.998 & 4.30 & 0.006 & 0.002 & 0.482\\ 
UF200 & 356.17 & 0.54 & 0.18 & 0.07 & 0.708 & 3.26 & 0.003 & 0.001 & 0.347\\
%\botrule
\hline
\end{tabularx}}
\label{table-PhyVol}
\end{table}

Control \emph{P. polycephalum} plasmodia exhibited stereotypical oscillatory electrical behaviour (Fig.~\ref{fig-Traces}). Mean period, mean potential and dominant frequency of all experimental plasmodia are summarised in Tab.~\ref{table-PhyVol}. The mean period of oscillation increased from 193 seconds to 232 and 356 with unfunctionalised 5~nm and 200~nm treatments, respectively. The results gained were variable and so statistical significance was not achieved (P=0.482 and P=0.347 respectively). Spiropyran-modified nanoparticles were not used to measure electrical properties because of the high fail rate of successful `wire' growth. %% Some repetition of table results here.

\begin{table}[!tbp]
\caption{Resistance measurements for \emph{P. polycephalum} and lettuce seedlings treated with gold nanoparticles. Adjusted resistance is an estimation of the `true' resistance obtained through subtracting the mean resistance of both agar islands in series (see Section \ref{sec-resultsResistance}). UF5/200: Unfunctionalised 5/200~nm varieties, SPM: spiropyran-modified.}
\begin{tabularx}{\textwidth}{@{}lccccr@{}}
\hline
\emph{P. polycephalum} & \multicolumn{1}{X}{Resistance (M$\Omega$s)} & \multicolumn{1}{X}{Adjusted Resistance (M$\Omega$s)} & SD & SE & P value\\
\hline
Control & 3.72 & 1.85 & 1.79 & 0.49 & --\\  
UF5 & 3.34 & 1.47 & 0.82 & 0.31 & 0.705\\ 
UF200 & 2.73 & 0.85 & 0.84 & 0.23 & 0.295\\
\hline
Lettuce seedlings & & & & & \\
\hline
Control & 4.81 & 2.94 & 1.52 & 0.38 & --\\
UF5 & 3.61 & 1.74 & 1.24 & 0.30 & 0.031\\
UF200 & 2.41 & 0.53 & 0.79 & 0.17 & $<$0.001\\
SPM & 3.76 & 1.89 & 1.29 & 0.37 & 0.119\\  
\hline
\end{tabularx}
\label{table-Res}
\end{table}

Lettuce voltage properties show an initial exponential decay followed by stabilisation; measurements of electrical potential were not averaged and were taken after 500 seconds when the voltage had stabilised. These results are summarised in Tab.~\ref{table-LetVol}. Mean electrical potential increased by 1~mV for both unfunctionalised 5~nm and 200~nm whereas increased by 2~mV for Spiropyran-modified. Statistical significance however was not achieved with unfunctionalised 5~nm (P=0.342), unfunctionalised 200~nm (P=0.460) or Spiropyran-modified (P=0.342).

\subsubsection{Resistance}
\label{sec-resultsResistance}

%% SHOULD PROBABLY NOTE SOMEWHERE WHY DOUBLE DISPENSATION VOLUME WAS USED WITH PHYSARUM
As agar islands themselves are resistors, the measurements detailed here are not true depictions of the organisms tested.  The resistance of the agar islands in isolation were measured by resting a 10~mm wire with known resistance of 
15.4$\Omega$ across the islands in a simulacrum the experiments for measuring the organisms' resistance: this value was utilised to calculate an approximation of the `true' resistance of the organisms using the resistor equation: 
$
R_{Total} = R_{\text{Agar Blob 1}} + R_{\text{Agar Blob 2}} + R_{\text{Wire}}
$
The resistance of \emph{P. polycephalum} tubes was also found to oscillate but at a much slower rate. The results of resistance measurements are summarised in Tab.~\ref{table-Res}. The resistance was shown to decrease in response to treatment, but by less than 1~M$\Omega$ in all tubes tested. The small change in resistance and variability of the data resulted in statistical significance not being achieved (P=0.705, P=0.295 for 5nm and 200nm respectively). Desiccated tubes were found to have infinite resistance. Resistance of lettuce seedlings treated with nanoparticles decreased by up to 2.5~M$\Omega$. Statistical significance was achieved with both unfunctionalised 5~nm and unfunctionalised 200~nm (P=0.031, P=$<$0.001) respectively. Whereas spiropyran-modified nanoparticles were not significantly different (P=0.119) even though there was a decrease in resistance of 1.1M$\Omega$.

\section{Discussion}
%% Slime looked healthy via histo!
% We mention that sp-mod particles are photochromic but we don't actually mention UV-ing them in methods!

\subsection{Observations and Toxicity}

All varieties of gold nanoparticle except unfunctionalised 5~nm were carried and internalised by \emph{P. polycephalum}, as was identified via SEM observations and EDX analyses. The method of internalisation is difficult to determine without using higher-resolution techniques, but all observations indicate that they were uptaken via vesicular endocytosis (in the same manner that \emph{P. polycephalum} consumes food substrates), as has been demonstrated for other nanoparticle varieties \cite{Mayne2015}. This provides an explanation for the apparent intracellular assembly of nanoparticles, i.e. aggregation following exposure to the intracellular environment. The absence of gold in histological sections is not necessarily an indication that gold was not present as the resolution of the technique was not adequate to visualise the deposits seen via SEM; the organism's histology appeared normal \cite{Mayne2011,Mayne2015,Kessler1982}.

Every variety of gold nanoparticle was also detected within lettuce seedlings via SEM/EDX, and black deposits suggestive of large gold deposits were found in histological sections. The lettuce seedlings appeared to have internalised significantly more gold than \emph{P. polycephalum} --- both via SEM and also due to its visible presence in histological sections --- perhaps indicating that it lacks a mechanism for preventing gold uptake, although (as aforementioned) it may also be considered to be a more resilient system. The mechanisms underlying the accumulation of environmentally-acquired metals in the epidermis is not abundantly clear, but it is nevertheless a previously-observed phenomenon \cite{Lavid2001}; nanoparticles were not observed in the xylem/phloem, but due to the fact that samples were dissected pre-fixation, diffusion of nanoparticles into the aqueous environment cannot be ruled out, hence no indications are available pertaining to their route of uptake.

Evidence of assembly of internalised particles was observed in both organisms, although neither size or coating appeared to have any bearing on the dimensions or morphology of aggregates. Coupled with the knowledge that size/coating did have some bearing on the electrical properties of the organism, however, (see Sect.~\ref{sec-elecprop}), this would seem to imply that size/coating affect the amount of suspension uptaken or/and that the differential intracellular distribution of particles is size-dependent to some degree. There was no statistically-relevant distinction between the size of internalised gold deposits between \emph{P. polycephalum} and lettuce seeds. 

The morphology of \emph{P. polycephalum} plasmodia treated with unfunctionalised nanoparticles was found to be somewhat emaciated; the effects of the buffers in this regard were discounted due to their low strength and previous indications from our laboratory indicating the organism's tolerance to them. Lettuce seedlings displayed no obvious reactions to exposure to unfunctionalised particles, but both organisms showed growth retardation, higher rates of propagation failure and morphological abnormality following exposure to the spiropyran-modified particles. This may, however, have been a result of exposure to their diglyme solvent, despite the low dispensation volumes used. Although the opinion of the nanotechnology community is somewhat split regarding the cytotoxicity of gold nanoparticles, these results lead us to believe that they are certainly not without their deleterious health effects in these model organisms.

\subsection{Electrical properties}
\label{sec-elecprop}
\subsubsection{\emph{P. polycephalum}}

Exposure to both varieties of unfunctionalised nanoparticle resulted in an increase in mean period and a corresponding decrease in dominant streaming frequency. This is possibly a result of alterations in plasmodial morphology. Although alterations in mean potential were observed, values were not statistically significant. \emph{P. polycephalum} `wire' resistance was found to decrease with both unfunctionalised treatments, although not to any degree of statistical significance. 

When plasmodial tubes were allowed to desiccate, their trails were found to be essentially non-conductive, hence eliminating the possibility of generating permanent wiring.

\subsubsection{Lettuce}

Conversely to the results presented for \emph{P. polycephalum}, treatment with 200~nm unfunctionalised nanoparticles resulted in an increase in mean potential by approximately 50\% (although not significant) and decrease in its resistance by 82\%. The 5~nm unfunctionalised variety also resulted in a statistically significant reduction in resistance, although by only 41\%. As such, the resistance of the organism was bought down to a value of much greater use in conventional circuitry, creating in effect a `living wire'. 

As with \emph{P. polycephalum}, the spiropyran-modified nanoparticles did not significantly alter the electrical properties of lettuce, despite the apparent abundance of gold in samples analysed via SEM/EDX, indicating that their coating may have been a factor in reducing any favourable characteristics --- to advance a somewhat more scientific hypothesis, it is possible that the spiropyran coating, even if irreversibly hydrolysed by contact with the cellular environment, blocks ionic wave conduction to some degree. 

% Put in actual numbers a p values\\
% Spiropyran is sometimes capitalised

\subsection{Conclusions}

The results detailed here suggest that gold nanoparticles are a suitable material for biohybridisation, but are not without their detriments: this is especially clear when comparing the apparent deleterious health effects suffered by \emph{P. polycephalum} to both unfunctionalised varieties of nanoparticle in biocompatible buffers. This is a somewhat surprising result when one considers the similarities between slime mould and mammalian cells, as the use of gold nanoparticle-containing products in novel and experimental medical therapies is already wide-spread \cite{Dykman2011,Huang2008,Copland2004}. Furthermore, the toxicological effects observed in both varieties of substrate used in this study augur badly regarding the environmental impact of gold nano-products in the event of accidental release. This further highlights that our knowledge of the interactions of nanomaterials formed from non-toxic (when macro-scale) materials with live substrates is still lacking. This does not limit the functionality of biohybridisation with gold nanoparticles for unconventional computing purposes, however, as the alterations in electrical properties observed are nevertheless substantial, despite the comparatively small quantities dispensed in each experiment.

The potential uses for this technology are manifold~\cite{adamatzky2012physarum}. The reduction of electrical resistance to K$\Omega$ range (when adjusted to remove the resistance of the agar island interfaces) allows for the generation of discrete macro-scale electronic components; the loss of photochromic properties of the spiropyran-modified particles when exposed to a cellular environment is unfortunate with regards to this particular application, although this is in no way indicative that other varieties of functionalised nanoparticle would be similarly affected. Another potential application is \emph{in vivo} nanoassembly to form discrete electrical components, e.g. capacitive elements, sensing units. The apparent alterations in size of gold aggregate in lettuce seedlings indicate that cytoplasmic assembly may be size-dependent to some degree and hence partially-controllable. Finally, the reduction in substrate resistance holds promise for enhancing the efficiency of computer-organism electrical interfaces. Previous studies with \emph{P. polycephalum} have indicated that the organism's electrical resistance effectively limits its usefulness in such applications \cite{Mayne2015,Mayne2015b, whiting2015transfer}, but the results presented here indicate that not only can both varieties of substrate be made significantly more conductive, but their other properties --- streaming frequency, electrical excitability etc. --- may also be manually altered. Enhancing the mean potential, for example, holds value in manual adjustment of baselines in electrically-coupled biological sensors.

%\section{Appendices}
%\subsection{Declaration of Interest}
%The authors declare no conflict of interest.

\section*{Acknowledgements}

This work was supported by the EU research project PHYCHIP: ``Physarum Chip: Growing Computers from Slime Mould'' (FP7 ICT Ref 316366)  and   SYMONE: ``Synaptic Molecular Networks for Bio-inspired Information Processing''  project
(FP7 ICY Ref 318597), both  FET-UCOMP thematic area, and in part by COST action MP1202 (HINT).

The authors extend their sincerest thanks to Dr. Ben De Lacy Costello, Dr. David Patton and Paul Kendrick of the University of the West of England for their invaluable support throughout this program of research.

\bibliography{references}
\bibliographystyle{elsarticle-num}

\end{document}